# PRELIMINARY ANALYSIS OF PERIODOGRAM SHAPES AND THEIR CLASSIFICATION

L.S. Kudashkina
Odessa National Maritime University,
Department of Mathematics, Physics and Astronomy
kuda2003@ukr.net

RV-type (RV Tau) stars are pulsating yellow supergiant stars whose light curves are characterized by alternating deep and shallow minima. They have spectral types F to G at maximum light, and K to M at minimum. The changes in the brightness are correlated with the changes of the spectrum. The period from one deep minimum to the next (the "formal" period) ranges from 30 to 150 days. The complete light amplitudes can reach 3-4 magnitudes in V. [1].

The RV-type stars are heterogeneous, and their evolutionary stage is not well understood. Often RV Tauri star may be misclassified as a Popular II Cepheid or a yellow semi-regular variable. Also most RV-type stars show expected decreases in the period, although (O-C) diagrams are dominated by random, cycle-to-cycle period fluctuations. Analysis of IRAS data show that the rate of mass loss seems to be decreasing, probably the stars go off with AGB.

Thus, RV-type stars are probably executing blue loops from the AGB, or are in their final transition from the AGB to the white dwarf stage. [2].

The position of the RV-type stars and related objects is shown on the figure 1.

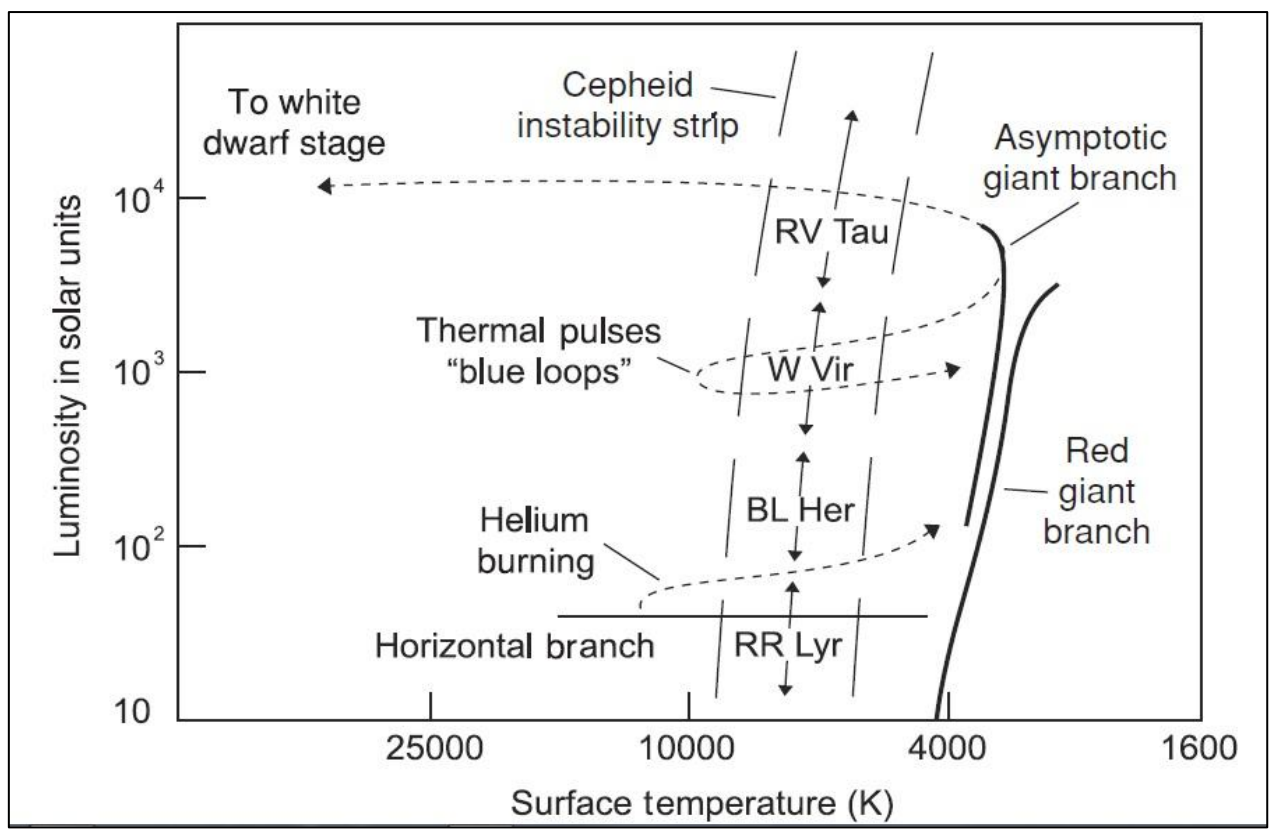

Fig 1. The position of the RV-type stars on HR-diagram. It's taken from the book of J.R. Percy [2].

A periodogram analysis of 47 stars noted in GCVS (1) as RV-type stars was carried out. Note that the research methods applied here to RV-type stars have previously been applied to long-period variables Mira-type stars and semi-variables. An overview of the results is given, for example, in the papers [3, 4]. For the analysis, the programs of I.L. Andronov and co-authors were used [5, 6, 7]. The technique of working with photometric observations of databases, such as AFOEV, AAVSO, ASAS and similar ones, has been repeatedly described in works [8, 9, 10]. To determine individual extrema of pulsating variables, we have used the software VSCalc [11] and, recently, the gradually improving versions of the software MAVKA [12, 13, 14]. The mean light curves folded with the best period and approximated using the program MCV [15].

Periodograms are previously classified by their shape, or rather the presence or absence of certain structures: two peaks in a 2:1 ratio, the presence of satellites at these peaks indicating the result of beats.

Table 1 shows the classification groups. Group I includes objects showing the periodogram typical form of RV Taurus stars, and the ratio of the periods of the two main peaks is indicated. Group II includes objects whose periodograms contain signs of multiperiodicity (Multi-p) or vice-versa, only one clear peak, instead of two (Single-p). Group III includes objects whose periodograms are highly noisy mainly due to the small number of observations. They do not show the typical details of RV-type stars.

The study identified stars that most likely do not belong to the RV Taurus type. Indications of these objects, as mistakenly classified, are available in other works, for example in [16].

Table 1. The classification groups according to periodogram shapes.

| I (19) | | II (7) | | III (17) | | Reclass (4) | |
|---|---|---|---|---|---|---|---|
| Star | P2/P1 | Star | | Star | | Star | Type |
| AC Her | 2.00 | MT Lyr | | TT Oph | | BI Cep | SR |
| AR Sgr | 2.00 | V 861 Aql | | TX Per | | OR Her | SR? |
| AZ Sgr | 1.99 | DY Aql | | V 381 Aql | | V 609 Oph | SR? |
| EQ Cas | 2.00 | SZ Mon | Single-p | AR Pup | | QV Aql | RV? |
| EZ Aql | 2.00 | V 967 Cyg | Single-p | RY Ara | | | |
| LN Aql | 2.00 | V 399 Cyg | Single-p | BT Lac | | | |
| RX Cap | 2.01 | AD Aql | Multi-p | SU Gem | ? | | |
| SS Gem | 2.00 | | | SX Cen | | | |
| TX Oph | 2.03 | | | DZ UMa | | | |
| UZ Oph | 1.95 | | | IW Car | | | |
| V Vul | 2.01 | | | GK Lac | | | |
| V 1008 Cyg | 2.66 | | | V 428 Aur | | | |
| V 453 Oph | 1.99 | | | V 590 Aql | | | |
| V 360 Cyg | 2.00 | | | AI Sco | Multi-p | | |
| U Mon | - | | | DF Cyg | Multi-p | | |
| EP Lyr | 2.00 | | | RV Tau | Multi-p | | |
| GK Cyg | 2.00 | | | TW Cam | Multi-p | | |
| V 457 Cyg | 2.01 | | | | | | |
| R Sge | 2.01 | | | | | | |

Figure 2 shows examples of periodograms typical of groups I, II, and III. On all periodograms, the frequency is set on the horizontal axis, and the value of the test function is set on the vertical axis.

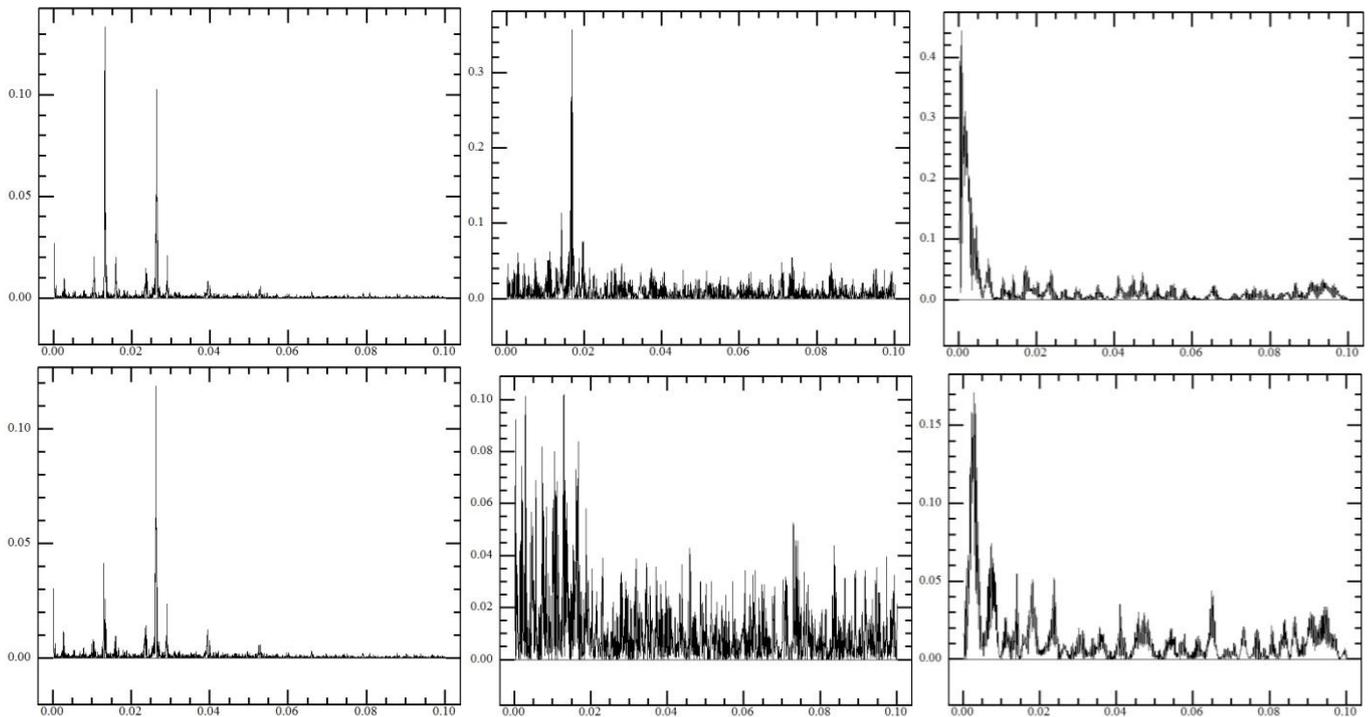

Fig. 2. The periodograms: V Vul – on the left (gr. I), V 399 Cyg – in the middle (gr. II), IW Car – on the right (gr. III). In the second periodograms (bottom) we have prewhitened the data by subtracting the contribution with a period corresponding to the highest peak.

About the reclassified objects, the following can be noted. In addition to those listed in table 1 in the corresponding column (Reclass), there are several other objects that do not clearly show signs of the behavior of stars of the RV-type. For example, the star SU Gem shows the variability characteristic of stars such as Mira Ceti. The peak on the periodogram corresponding to the period P=684.3±0.5 has the greatest value of the test function. In GCVS [1] close to this value is given as the period of change in the average brightness. However, the period of 50 days specified in the same place was not revealed according to our researches. According to the observations of different databases, the period of 684 days is specified by the FDCN (FOUR-N) program [17] in the range from 682 to 715 days. A summary of the results is given in table 2. Only the spectral type F5-M3 indicates a star hotter than Mira. This star can be classified as a semi-regular variable.

In all the figures of the light curves, the phase is set on the horizontal axis, and the brightness of the star is set on the vertical axis.

Table 2.

| Databases | The elements was calculated by FDCN | Number of obs. |
|---|---|---|
| AFOEV | $JD_{max}= (2451932\pm4) + (683.9\pm 0.4)^{d}E$ | 1994 |
| AASVO-vis | $JD_{max}= (2451907.1\pm 1.9) + (682.3\pm 0.3)^{d}E$ | 2327 |
| AASVO-V | $JD_{max}= (2454631\pm7) + (685.3\pm 0.4)^{d}E$ | 646 |
| AASVO-B | $JD_{max}= (2449848\pm98) + (686.2\pm0.6)^{d}E$ : | 51 |
| ASAS-V | $JD_{max}= (2457361\pm4) + (714\pm4)^{d}E$ : | 94 |

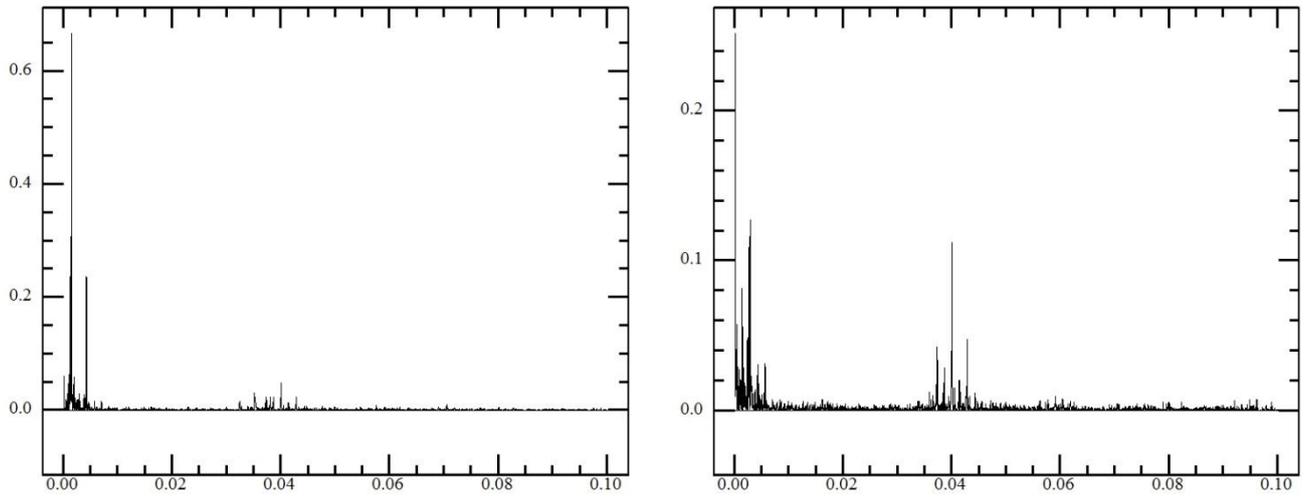

Fig. 3. Periodograms for the star SU Gem. The high peak corresponds to a period of 684.3 days. The right periodogram is obtained after subtracting the contribution with a period of $684.3^d$.

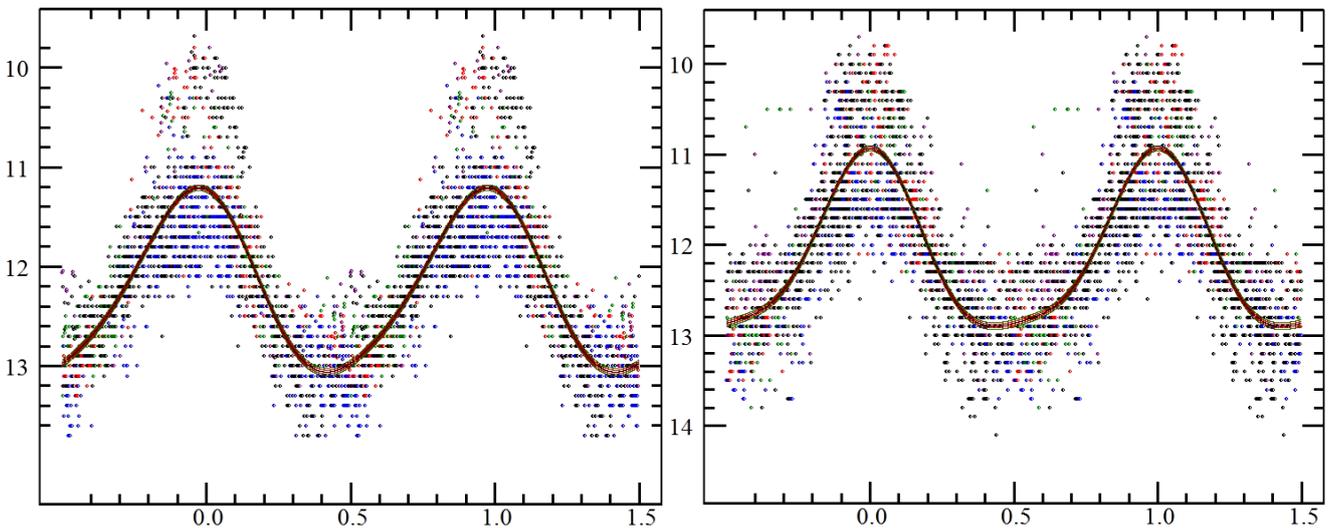

Fig. 4. The mean light curve of SU Gem with a period of $684^d$, obtained from observations of members AFOEV (left) and from observations of members AAVSO (right).

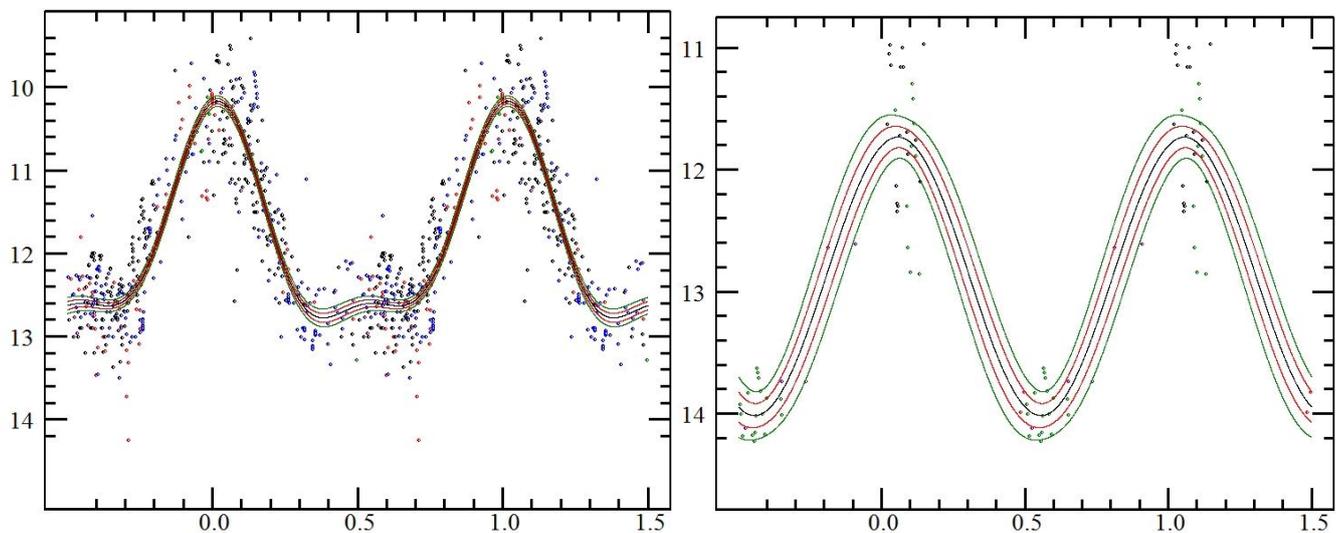

Fig. 5. The mean light curve of SU Gem with a period of $685^d$, obtained from observations of members AAVSO in V- (left) and B-band with a period of $686^d$ (right).

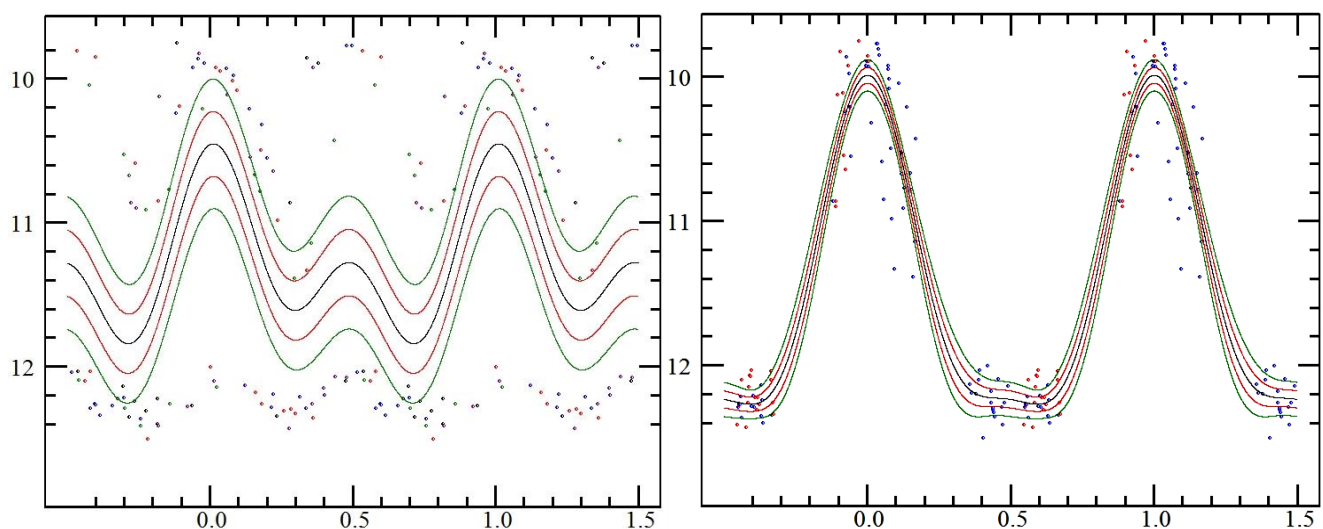

Fig. 6. The mean light curve of SU Gem, obtained from observations of ASAS with period $50^d$ (left) and with period and with a period that has been refined to days $715^d$ (right).

This study is done in the framework of projects "Inter-Longitude Astronomy" [18, 19, 20], "Ukrainian Virtual Observatory" [21] and "Astroinformatics" [22].

Acknowledgement. The authors are thankful to Kavka S. [23] and the members of the AAVSO for their observations and to Prof. I.L. Andronov for useful discussions.